\documentclass[12pt]{article}
\usepackage{graphics,color}
\begin{document}
\thispagestyle{empty}
\noindent\
\\
\\
\\
\begin{center}
\large \bf Excited weak bosons and their decays
\end{center}
\hfill
 \vspace*{1cm}
\noindent
\begin{center}
{\bf Harald Fritzsch}\\
Department f\"ur Physik\\ 
Ludwig-Maximilians-Universit\"at\\
M\"unchen, Germany \\
\end{center}

\begin{abstract}

The weak bosons are not elementary gauge bosons, but bound states of two fermions. Here the excitations of the weak bosons are discussed. Especially we study the decays of these excited states into weak bosons and photons. 

\end{abstract}

\newpage

The weak bosons might not be elementary gauge bosons, but bound states of two fermions, analogous to the $\rho$-mesons in QCD. The weak bosons are the ground states. The scalar boson with a mass of 125 GeV, discovered at the LHC (ref. 1,2), would be an excitation of the $Z$-boson.\\

A theory, in which the weak bosons are bound states, describes also the leptons and quarks as bound states. Such theories were studied in the past by many theorists (ref. 3,...,11). An unsolved problem for such theories is to explain, why the leptons and quarks, except the top-quark, have masses, which are much smaller than the masses of the weak bosons. \\ 

The weak bosons are bound states of a fermion and its antiparticle, which are denoted as "haplons" (see also ref. 7 and 9). Their dynamics is described by a confining gauge theory, denoted as "quantum haplodynamics" ($QHD$).\\  

The $QHD$ mass scale is given by a mass parameter $\Lambda_h$, which determines the size of the weak bosons. The $QHD$ mass scale is about 0.2 TeV (ref. 4,5). The haplons are massless and interact with each other through the exchange of massless gauge bosons.\\
 
Two types of haplons are needed as constituents of the weak bosons, denoted by $\alpha$ and $\beta$. The charges of the weak bosons do not fix the charges of the haplons. They could be, for example, (+1/2) and (-1/2), but they might also be the same as the charges of the quarks: (+2/3) and (-1/3). We shall assume, that this is the case.\\ 

The three weak bosons have the following internal structure:
\begin{eqnarray}
W^+ = (\overline{\beta} \alpha),~~ W^- = (\overline{\alpha} \beta),~~W^3 =\frac{1}{\sqrt{2}} \left( \overline{\alpha} \alpha - \overline{\beta} \beta \right).
\end{eqnarray}

The weak bosons consist of pairs of haplons, which are in an s-wave. The spins of the two 
haplons are aligned, as the spins of the quarks in a $\rho$-meson. The first excited states are those, in which the two haplons are in a p-wave. The weak isospin and the angular momentum of these states are described by the two numbers $(I,J)$.\\  

There are three $SU(2)$ singlets: S(0), S(1) and S(2) as well as three $SU(2)$ triplets: T(0), T(1) and T(2). These bosons are analogous to the mesons in strong interaction physics, in which the quarks are in a p-wave. The scalar meson $\sigma$, the vector meson $h_1(1170)$ and the tensor meson $f_2(1270)$ are the  $QCD$-analogs of the singlet states $S(0)$, $S(1)$ and $S(2)$.\\ 

The isospin triplet mesons, the scalar meson $a_0(980)$, the vector meson $b_1(1235)$ and the tensor meson $a_2(1320)$, correspond to the bosons $T(0)$, $T(1)$ and $T(2)$.\\

We assume that the boson $S(0)$ is the particle, discovered at CERN - thus the mass of $S(0)$ is about 125 GeV. In analogy to QCD we expect that the masses of the other p-wave states are between 0.26 TeV and 0.41 TeV. Here are the expected masses of the bosons $S(1)$ and $S(2)$:
\begin{eqnarray}
S(1): 0.32~ TeV \pm 0.06~ TeV,\nonumber\\
S(2): 0.34~ TeV \pm 0.06~ TeV.
\end{eqnarray}

The masses of the $SU(2)$-triplet bosons $T$ are slightly larger than the masses of the $S$-bosons: 
\begin{eqnarray}
T(0): 0.25~ TeV \pm 0.05~ TeV,\nonumber\\
T(1): 0.33~ TeV \pm 0.05~ TeV,\nonumber\\
T(2): 0.36~ TeV \pm 0.06~ TeV.
\end{eqnarray}

The $S(0)$-boson will decay into two charged weak bosons, into two $Z$-bosons, into a photon and a $Z$-boson, into two photons, into a lepton and an anti-lepton or into a quark and an anti-quark. Details were discussed in ref. 5. The expected decay rates are similar to the decay rates, expected for a Higgs-particle.\\

According to the theory a Higgs particle decays into two fermions with the probability of about 66.7 \%, into two gluons with the probability of about 8,6 \% and into two gauge bosons with the probability of about 24,7 \%. In our model there is no decay of the $S(0)$-boson into two gluons. Decays into pairs of leptons and quarks are possible, but cannot be calculated.\\ 
 
The bosons $S(1)$ and $S(2)$ and the nine $T$ - bosons will decay mainly into two or three weak bosons or photons and into a pair of leptons or quarks. The bosons $S(1)$ and $T(1)$ are vector bosons, thus they cannot decay into two photons. I shall consider in detail the decays of the charged boson $T(0,+)$ and of the neutral boson $T(0,0)$, taking into account the available phase space.\\

The decay strengths for the decays cannot be calculated, since they depend on details of the $QHD$ dynamics. We parametrize the decays into two weak bosons or photons by a parameter A and the decays into three weak bosons or photons by a parameter B, taking into account the phase space. We expect that the parameters A and B are about equal. Here are the branching ratios for the decays of 
$T(0,0)$ into two bosons:
\begin{eqnarray}
&&BR~(T(0,0)=> W^+ + W^-) \approx (1/3)\cdot A \cdot \left(1-\frac{M^2_W}{M^2_T}\right)^4 \;,\nonumber \\
&&BR~(T(0,0)=> Z + Z) \approx 0.197 \cdot A \cdot \left(1-\frac{M^2_Z}{M^2_T}\right)^4 \;,
\nonumber \\
&&BR~(T(0,0)=> Z + \gamma) \approx 0.119 \cdot A \cdot \left(1-\frac{M^2_Z}{M^2_T}\right)^2 \;,\nonumber \\
&&BR~(T(0,0)=> \gamma + \gamma)\approx 0.018 \cdot A.
\end{eqnarray}\\

For the decays into three particles we obtain: 
\begin{eqnarray}
&&BR~(T(0,0)=> W^+ + W^- + \gamma)\approx (2/3)\cdot 0.231 \cdot B \cdot \left(1-\frac{M^2_W}{M^2_T}\right)^4 \;,\nonumber \\
&&BR~(T(0,0)=> W^+ + W^- + Z)\approx (2/3)\cdot 0.769 \cdot B \cdot \left(1-\frac{M^2_W}{M^2_T}\right)^4 \left(1-\frac{M^2_Z}{M^2_T}\right)^2\;,\nonumber \\
&&BR~(T(0,0)=> Z + Z + Z)\approx (1/3) \cdot B \cdot 0.454 \cdot \left(1-\frac{M^2_Z}{M^2_T}\right)^6 \;,\nonumber \\
&&BR~(T(0,0)=> Z + Z + \gamma) \approx (1/3) \cdot B \cdot 0.410 \cdot \left(1-\frac{M^2_Z}{M^2_T}\right)^4 \;,\nonumber \\
&&BR~(T(0,0)=> Z + \gamma + \gamma) \approx (1/3) \cdot B \cdot 0.123 \cdot \left(1-\frac{M^2_Z}{M^2_T}\right)^2 \;,\nonumber \\
&&BR~(T(0,0)=> \gamma + \gamma + \gamma)\approx (1/3)\cdot B\cdot 0.012.
\end{eqnarray} \\

Here we have assumed, that the $T(0,0)$ does not decay into pairs of leptons or quarks. Otherewise the branching ratios would be smaller. All branching ratios have to be multiplied with a parameter r. For the $S(0)$-boson one finds r $\approx 0.25$.\\ 

If we assume, that the mass of $T(0,0)$ is 0.25 TeV, that the parameters A and B are equal and that there is no decay into two leptons or quarks, we obtain the following branching ratios:\\
\begin{eqnarray}
&&BR~(T(0,0)=> W^+ + W^-) \approx 0.37 \;,\nonumber \\
&&BR~(T(0,0)=> Z + Z) \approx 0.09 \;,\nonumber \\
&&BR~(T(0,0)=> Z + \gamma) \approx 0.08 \;,\nonumber \\
&&BR~(T(0,0)=> \gamma + \gamma)\approx 0.02 \;,\nonumber \\
&&BR~(T(0,0)=> W^+ + W^- + \gamma)\approx 0.08 \;,\nonumber \\
&&BR~(T(0,0)=> W^+ + W^- + Z)\approx 0.21 \;,\nonumber \\
&&BR~(T(0,0)=> Z + Z + Z)\approx 0.05 \;,\nonumber \\
&&BR~(T(0,0)=> Z + Z + \gamma) \approx 0.07 \;,\nonumber \\
&&BR~(T(0,0)=> Z + \gamma + \gamma) \approx 0.03.
\end{eqnarray}\\

In a similar way we calculate the decay rates for the decay of the charged boson $T(0,+)$. If the mass of this boson is 0.25 TeV and if the parameters A and B are equal, we obtain for the branching ratios: 

\begin{eqnarray}
&&BR~(T(0,+)=> W^+ + Z) \approx 0.40 \;,\nonumber \\
&&BR~(T(0,+)=> W^+ \gamma) \approx 0.15 \;,\nonumber \\
&&BR~(T(0,+)=> W^+ + W^+ + W^-)\approx 0.30 \;,\nonumber \\
&&BR~(T(0,+)=> W^+ + Z + Z)\approx 0.08 \;,\nonumber \\
&&BR~(T(0,+)=> W^+ + Z + \gamma)\approx 0.06 \;,\nonumber \\
&&BR~(T(0,+)=> W^+ +\gamma + \gamma)\approx 0.01.
\end{eqnarray}

The excited weak bosons can be observed in the LHC experiments. The neutral boson $T(0,0)$ decays into two photons with a branching ratio of about 2 \%, which is similar to the branching ratio for the two-photon decay of the boson $S(0,0)$. Thus the neutral boson might be discovered in the two-photon decay.\\ 

If the $T(0,0)$ boson has a mass of 0.25 TeV, the expected rate for the decay into two photons should be about 0.08 times the rate of the decay of the $S(0)$ boson into two photons. The reduction is due to the higher mass of the $T(0,0)$ boson (see also ref. 15).\\

The $T(0,0)$ boson might also be observed in the decay into two charged weak bosons (branching ratio about 37 \%), in the decay into two Z-bosons (branching ratio about 9 \%) or in the decay into a Z-boson and a photon (branching ratio about 8 \%).\\

The charged bosons $T(0,+)$ and $T(0,-)$ can be observed by the decay into a charged weak boson and a photon (branching ratio about 16 \%) or into a charged weak boson and a Z-boson (branching ratio about 40 \%). Decays into three particles, e.g. the decay into a charged weak boson and two Z-bosons, have small branching ratios and a very difficult to observe.\\

If our model is correct, the Large Hadron Collider should soon discover the new boson $T(0,0)$, e.g. by observing the decay of this particle into two photons, and the new charged bosons $T(0,+)$ and $T(0,-)$, e.g. by observing the decay into a charged weak boson and a photon. Afterwards the bosons $S(1)$ and $S(2)$ will be discovered, e.g. by observing the decay of the boson $S(1)$ into two Z-bosons and the decay of the boson $S(2)$ into two photons.\\ 

Then the three bosons $T(1)$ might be observed, e.g. by the decay into two Z-bosons or into a Z-boson and a charged weak boson. To observe the bosons $T(2,+)$,  $T(2,0)$ and $T(2,-)$ is very difficult, since the branching ratios for the decays into two photons or a photon and a charged weak boson are very small. The $T(2)$-bosons will mainly decay into three or four particles and it will take a long time, until the bosons $T(2)$ are discovered.\\

\end{document}